\documentclass[mypaper,7pt,twoside]{CoAst}
\usepackage{epsf,graphicx,fancyhdr}
\renewcommand{\chaptermark}[1]{\markboth{#1}{}}

\newcommand{\kopf}{\small\itshape Comm. in Asteroseismology \\ Contribution to the Proceedings of the 38$^{th}$\,LIAC\,/\,HELAS-ESTA\,/\,BAG, 2008
}

\newcommand{\Authors}[1]{\begin{center}\normalsize\bf\sf #1 \end{center}}

\renewcommand{\author}[1]{\begin{center}\normalsize\bf\sf #1 \end{center}}
\newcommand{\Address}[1]{\begin{center}\small\sf #1 \end{center}}

\DeclareGraphicsExtensions{.eps,.jpg}

\newcommand{\Session}[1]{{\vspace{3mm}\small \noindent  \hspace*{3mm} Session: } #1 \normalsize}

\newcommand{\Objects}[1]{{\vspace{0mm}\small \noindent  \hspace*{3mm} Individual Objects: } \small #1 \normalsize}

	\newcommand{\three}{\small Atmospheres, mass loss and stellar winds}

\renewenvironment{abstract}{\section*{Abstract}\normalsize\sf}{}
\newcommand{\References}[1]{\begin{flushleft}{\large References\\}\vspace*{2mm}\small #1 \end{flushleft}}

\newcommand{\chapterCoAst}[2]{\chapter[\sf\normalsize #1\\ \footnotesize \hspace*{5mm}by #2 \sf\normalsize][]{#1\\}\rhead[\fancyplain{}{\sf\footnotesize \center{#1}}]{\fancyplain{}{\sffamily\thepage}}\lhead[\fancyplain{\kopf}{\sffamily\thepage}]{\fancyplain{\kopf}{\sf\footnotesize \center{#2}}}}

\newcommand{\figureCoAst}[5]{\begin{figure}[#4]
\centering
\includegraphics*[#5]{#1}
\caption{#2}
\label{#3}
\end{figure}}

\renewcommand{\chaptername}{}

\def\rfr{\smallskip\par\noindent
        \hangindent=7truemm
        \hangafter=1}

\begin{document}
\sf

\chapterCoAst{Non-Radial Pulsations and Large-Scale Structure in 
              Stellar Winds}
{R.\,Blomme} 
\Authors{R.\,Blomme} 
\Address{Royal Observatory of Belgium \\
         Ringlaan 3, B-1180 Brussel, Belgium
}

\noindent
\begin{abstract}
Almost all early-type stars show Discrete Absorption Components (DACs)
in their ultraviolet spectral lines. These can be attributed to
Co-rotating Interaction Regions (CIRs): large-scale spiral-shaped
structures that sweep through the stellar wind.

We used the Zeus hydrodynamical code to model the CIRs. In the model,
the CIRs are caused by ``spots" on the stellar surface. Through the
radiative acceleration these spots create fast streams in the stellar
wind material. Where the fast and slow streams collide, a CIR is
formed. By varying the parameters of the spots, we quantitatively fit
the observed DACs in HD~64760.

An important result from our work is that the spots do not rotate
with the same velocity as the stellar surface. The fact that the
cause of the CIRs is not fixed on the surface eliminates many potential
explanations. The only remaining explanation is that the CIRs are due
to the interference pattern of a number of non-radial pulsations.
\end{abstract}

\Session{ \three } 
\Objects{HD\,64760, $\zeta$\,Pup, $\xi$\,Per}

\section*{Structure in O-type stellar winds}

\figureCoAst{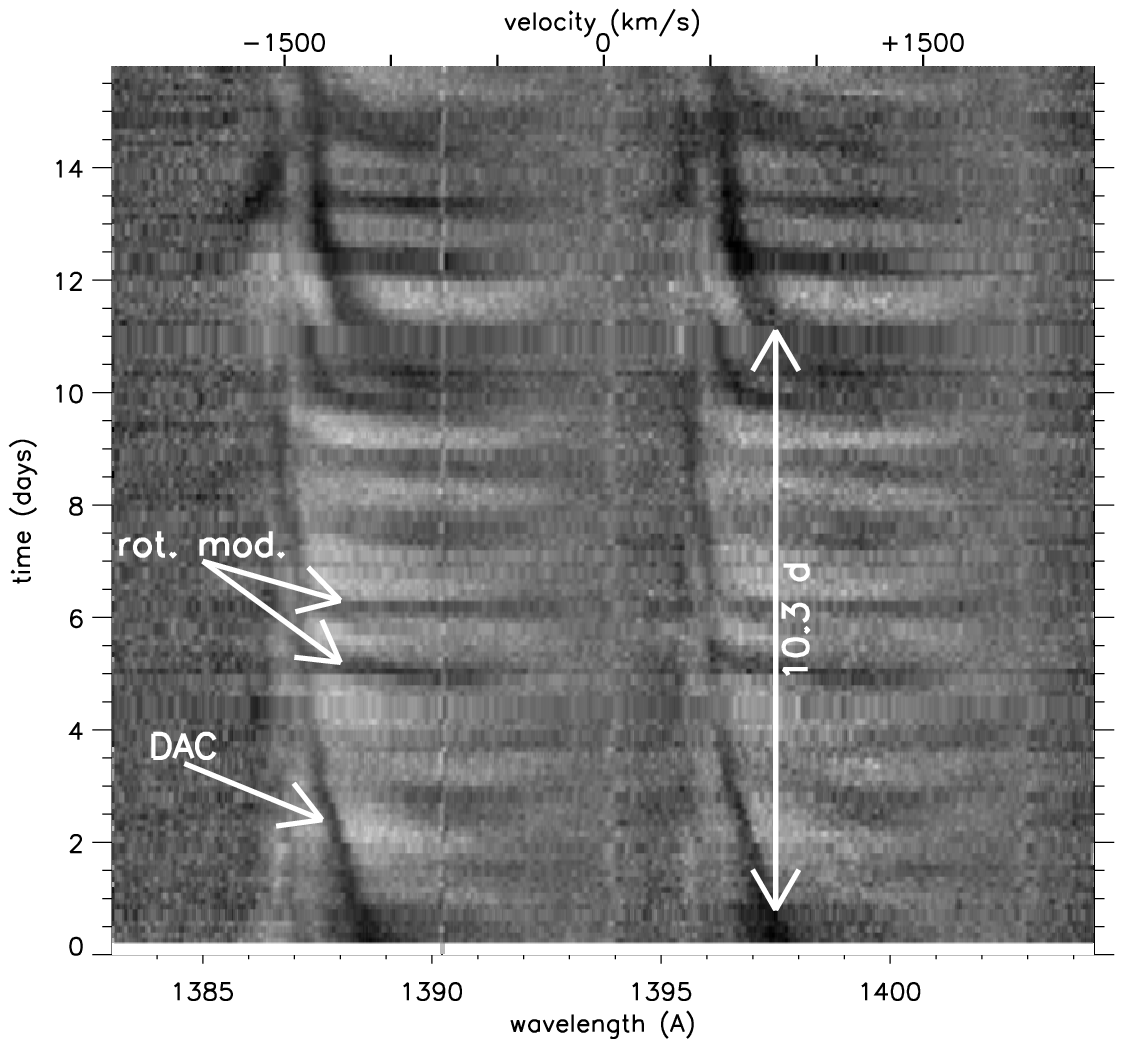}
{Grey-scale plot of the Si IV $\lambda\lambda 1394,1403$ P Cygni profile
of HD~64760,
based on spectra extracted from the IUE archive 
(observers: Massa et al.~1995).
At each wavelength the grey value indicates the difference
between this 
spectrum and the average spectrum. The spectra
are stacked as a function of time. DACs are clearly seen as the absorption
that is slowly moving toward 
the terminal velocity ($v_\infty = 1500$~km\,s$^{-1}$). 
The recurrence time
of the DACs is $\sim$~10.3~d. The rotational modulations move
much faster than the DACs, and repeat every $\sim$~1.2~d.
}
{grey-scale}{t}{clip,angle=0,width=115mm,height=9.1cm}

Many indicators exist for the presence of structure in the stellar winds
of early-type stars. Here we discuss 
the wind
structure in O and early B-type stars;
structure in Wolf-Rayet stars is discussed by Gosset (these proceedings).
Two types of structure are known to co-exist
in the stellar wind: 
small-scale and large-scale. The small-scale type (clumping)
has been discussed by Puls (these proceedings); 
we will concentrate on
the large-scale structure. The best-known indicators of 
this large-scale structure
are the Discrete Absorption Components (DACs). These can be seen
in the P Cygni profiles of some ultraviolet resonance lines.
The absorption part of such a P Cygni profile covers the whole
range of velocities from the base of the wind up to the terminal velocity
at large distances from the star. The DACs 
are additional absorption features that are 
superposed on these P Cygni profiles.
With time, the DACs move from 
low outflow velocities to
the terminal velocity. 

This behaviour is best seen
on a so-called dynamic spectrum (Figure~\ref{grey-scale}), 
which is a grey-scale plot that shows how much a single spectrum differs
from the average of all spectra. 
(Note that other authors use the highest fluxes as a reference spectrum, 
rather than the average. This explains why Figure 1 shows both emission
and absorption.)
By stacking the spectra in time-sequence,
one obtains a clear view of the outward movement of the DAC.
From the DAC behaviour of a sample of OB stars, it was found that
the DACs recur on timescales related to $v_{\rm eq} \sin i$
(Henrichs et al.~1988, Prinja 1988).
In addition to DACs, Figure~\ref{grey-scale} also shows rotational modulations:
these are the horizontal, nearly-straight features seen in the spectrum.
They cover a 
large range of velocities and (in this example) repeat on a $\sim$~1.2 day
timescale.
DACs have been detected in almost all O and early B-type stars
(e.g. Kaper et al.~1996, 1997, 1999),
but rotational modulations are much rarer.

The effect of structure is also seen in the H$\alpha$ line, which is
partly formed in the stellar wind of these early-type stars. Examples
of this are given by, e.g., Morel et al.~(2004).

\section*{HD~64760}
\subsection*{Rotational Modulations}

HD~64760 is a B0.5 Ib supergiant with 
$v_{\rm eq} \sin i$ = 265 km\,s$^{-1}$ (Kaufer et al.~2006), 
which is quite high
for a supergiant. The star 
is therefore probably viewed 
nearly equator-on, and, in what follows we will assume that it 
is viewed
exactly equator-on 
as this will simplify the analysis.

Figure~\ref{grey-scale} shows the data from the IUE Mega Campaign 
(Massa et al.~1995) for this star.
The plot is limited to the Si IV $\lambda\lambda 1394,1403$ doublet.
All features are seen in both components of the 
doublet, but we will concentrate our discussion on the 
(strongest) blue component to avoid contamination effects of the red
component.
A Fourier analysis shows that the rotational modulations are not exactly
straight, but bowed. This inspired Fullerton et al.~(1997) to propose a
simple ad-hoc model where a spiral-shaped structure is assumed 
to rotate through the wind. As this spiral
crosses the line of sight 
towards the observer, additional absorption is seen which starts
at an intermediate velocity and then moves to both higher and lower velocities
with time, thus creating a bow-shaped rotational modulation in the
spectrum.

A more theoretical background to this ad-hoc model was provided by
Cranmer \& Owocki (1996) 
who followed up on the idea proposed by Mullan (1986) that
Co-rotating Interaction Regions (CIRs) should exist in OB star winds.
They used a numerical code to solve the hydrodynamical equations of the 
stellar wind including the line driving, which is responsible for these
radiatively driven winds.
They then added a bright spot on the stellar surface 
centred on the equator
and calculated the effect 
of the additional line driving due to this spot. 
Because of the difference in radiative acceleration, the wind 
streaming out above the
bright spot has a higher density and a lower velocity than the wind outside
the spot. Due to the rotation, these streams collide and form a spiral-shaped
density pattern, called a Co-rotating Interaction Region (CIR).
It is important to realize that the CIR is a pattern in 
the wind; it is {\em not} the path the particles follow through the wind. 
Particles 
basically move out radially (except close to the stellar surface where 
some effect of the rotation is still seen, due to
angular momentum conservation).

\figureCoAst{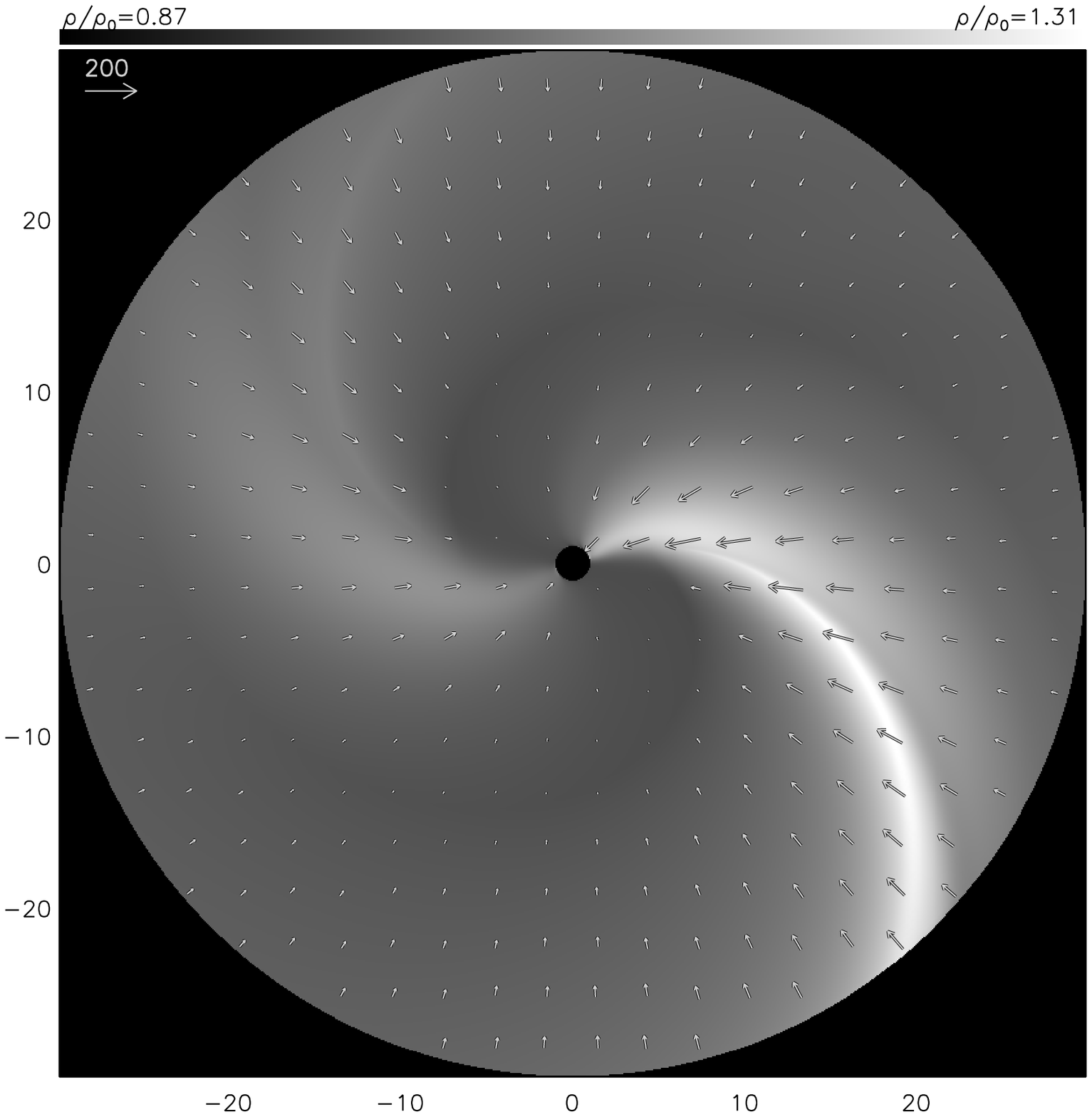}
{Model CIRs in the stellar wind of HD~64760. The grey-scale indicates the
density contrast (with respect to a smooth wind) and the velocity vectors
the deviation from the smooth wind velocity. The spots causing the
CIRs have a brightness enhancement of 20 \% and 8 \% and an opening
angle of 20$^{\rm o}$ and 30$^{\rm o}$, respectively. 
Both spots rotate at 1/5 of the equatorial velocity.
This model gives the 
best fit to the observed HD~64760 DACs seen in the Si IV 
line.
}
{CIR}{t}{clip,angle=0,width=91mm}

\subsection*{DACs}

Inspired by the above success for rotational modulations, 
Lobel and Blomme (2008) tried to explain the DACs
in a similar way. 
We calculated models using the Zeus hydrodynamics code, following the
procedure outlined by Cranmer \& Owocki~(1996).
A major difference in our work, however, is that
we included a spot that is not fixed on the 
stellar 
surface (an idea which had already been proposed by Kaufer et al. 2006).
The reason for introducing such a spot
is that the recurrence timescale 
of the DACs is 10.3~d (see Figure~\ref{grey-scale}), while the rotation period 
of HD~64760 is only 4.1~d. 
The main effect of a spot velocity different from the rotational velocity is 
that the CIRs spiral more for a rapidly rotating spot and less for a slower 
spot.
An example of the resulting hydrodynamical structure is shown in
Figure~\ref{CIR}.

For a given hydrodynamical model, we calculate the P Cygni profiles 
using the 3D radiative transfer code Wind3D developed by 
A. Lobel (see Lobel \& Blomme 2008). 
In this way, we can create artificial
grey-scale plots and compare their DACs to the observed ones.
By varying the parameters of the 2 spots we put in the model (brightness
enhancement and opening angle 
of the spot), we obtained a very good fit to the
shape and width of the DACs observed in the HD~64760 Si IV resonance
line (Lobel \& Blomme 2008).

As has already been pointed out by Cranmer \&
Owocki (1996), the velocity plateaus formed due to the CIRs are more important 
than the density enhancements in determining the effect on the P Cygni 
profiles. This can easily be seen from an extreme simplification of the 
radiative transfer, where we just consider the Sobolev approximation
to the optical depth ($\tau$) and limit ourselves to the radial direction only. 
We then have:
\begin{equation}
\tau \propto \frac{\rho}{{\mathrm d}v/{\mathrm d}r},
\end{equation}
from which it follows that both density ($\rho$) and 
velocity gradient (${\mathrm d}v/{\mathrm d}r$) can 
provide additional absorption. 
By plotting these variables for our set of models, we found that
in most cases it is the 
nearly-zero
velocity gradient that is responsible 
for the DACs.

The density contrast we find for the best-fit CIRs
is quite small (20 to 30~\%) 
and the total mass loss is increased by less than 1 \%
compared to a smooth wind.
These CIR structures are therefore quite delicate, and it is 
surprising that they are not destroyed by the instability mechanism
that is responsible for the clumping in the wind.
Owocki (1998)
has made some calculations combining CIRs and the instability mechanism.
He found that,
depending on the strength of the instability mechanism, the CIRs can be 
completely destroyed. 
However, observations tell us that DACs are seen in nearly all O and
early B-type stars
(Prinja \& Howarth 1986).
In principle, we can therefore use the presence of DACs to constrain
the amount of clumping that can be present in a stellar wind.

\subsection*{Non-Radial Pulsations (NRPs)}

The fact that 
the CIRs rotate more slowly than
the stellar surface eliminates many potential
explanations. 
A slower rotation timescale would be possible in a differentially 
rotating star, if the spots were located at higher latitude. However,
in that case, the nearly-radial outflow of the wind would prevent the 
CIRs from crossing in front of the stellar disc, and would therefore 
not explain the DACs.
The only remaining explanation is that the CIRs are due
to NRPs (or a beat pattern of NRPs).
From the spectra of photospheric lines, 
Kaufer et al.~(2006) found NRPs in HD~64760 with three different periods
($P_1$=4.810~h, $P_2$=4.672~h, $P_3$=4.967~h). 
They also found variability in the H$\alpha$ profile
(a line which is formed partly in the stellar wind), but with
a 2.4-day period. They tried to attribute this period to a beat period
between $P_1$ and $P_2$, but found that this beat period is 6.8~d
rather than 2.4~d.

However, if we look at the power 
spectrum they published (their Figure 5), we see that the NRP periods are not 
that well determined. A small shift of the $P_1$ and $P_2$
periods (by only 0.0235 hrs, which is well within the uncertainty) would
result in a beat pattern of 10.3~d, which is exactly
the recurrence timescale of the 
DACs.
It would be very interesting to see if the NRP periods of HD~64760 could be 
determined to such a precision
that the beat period can indeed be shown to correspond with the DAC recurrence
time.

\subsection*{Rotational Modulations revisited}

The claim that rotational modulations are well explained by CIRs is based
on a kinematical model (Owocki et al.~1995), i.e. a model that uses a 
pre-specified density wave superposed on a smooth wind.
When we try to make a hydrodynamical 
spot
model, however, 
we never achieve a rotational modulation that is sufficiently flat
to compare well with the observations (see Figure~\ref{rotmod}). 

The curvature of the rotational
modulation is mainly determined by the ratio $v_{\rm eq}/v_\infty$.
Introducing a spot with a velocity lower than $v_{\rm eq}$
might suggest itself 
as a possible solution, but in that case the timescale needs to
be stretched (because the spot takes a longer time to cross 
the line of sight as it moves 
more slowly). 
This stretching basically compensates for the effect of the slower spot velocity
and we again end up with a curvature incompatible with the observed
one. We thus have considerable problems in finding a
{\em hydrodynamical} model 
based on surface spots that can explain
the rotational modulations.

\figureCoAst{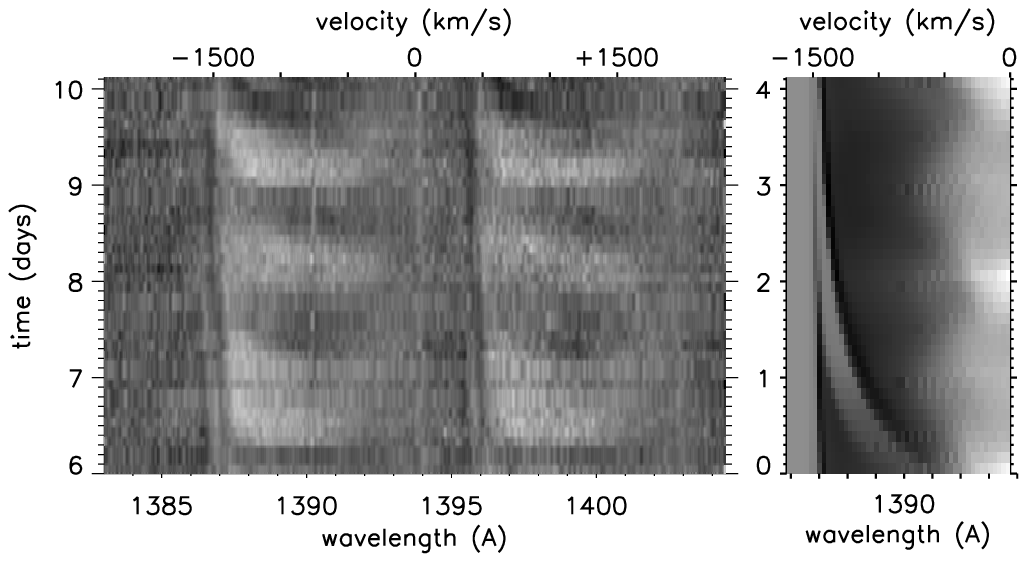}
{Attempted hydrodynamical 
spot
model ({\em right}) to explain the observed
rotational modulations ({\em left}). It is clear that the curvature of
the model (toward the terminal velocity) is much too high to provide
a good explanation for the observation. None of the models we tried
gave a satisfactory fit to the data.
}
{rotmod}{t}{clip,angle=0,width=115mm}

\section*{Other stars}

We also started modelling the O7.5III(n)((f)) star
$\xi$~Per, which has an earlier spectral 
classification than HD~64760. 
For this star it is not clear if the spots responsible for
the DACs are fixed on the stellar surface or not. The rotational period 
is less well determined (there is some discussion in the literature about 
the radius of this star) and it is possible that the DAC recurrence time
scale of 2.1 or 4.2 day is compatible with the (uncertain) rotation period of 
3.1 day. An interesting feature of
$\xi$~Per is that it sometimes shows two DACs at the same time that
(apparently) cross one another. We have succeeded in finding theoretical
models that also show such crossing DACs. It should be stressed that 
the crossing of the DACs is 
only
a line-of-sight effect; the corresponding
CIRs do not physically cross one another.

Another well-known star is $\zeta$~Pup (spectral type O4I(n)f), 
which has both DACs and rotational
modulations. Compared to HD~64760, the $\zeta$~Pup rotational modulations
are more slanted, straight lines
(Howarth et al.~1995).
The complete absence of curvature will present a further challenge 
to hydrodynamical modelling
on top of those already presented by HD~64760.

\section*{Conclusions}

Different types of structure are present in a stellar wind. CIRs are
one form of large-scale structure, which can very well explain the DACs,
but not the rotational modulations (contrary to what had been previously 
thought). Explaining the rotational modulations therefore remains a
challenge for the hydrodynamical modelling.
Specifically for HD~64760, the CIRs are related to a beat 
pattern of NRPs.
We caution, however, that it is not certain if we can extrapolate this 
conclusion to other stars, such as $\xi$~Per, where a magnetic field
might explain the observations as well.
The additional mass loss rate required 
for 
the CIRs is very small. It is surprising 
therefore that 
these structures
are not destroyed by clumping. This fact can, in principle, be used to 
constrain the amount of clumping in the stellar winds of early-type stars.


\References{
\rfr Cranmer, S.R., \& Owocki, S.P. 1996, ApJ, 462, 469
\rfr Fullerton, A.W., Massa, D.L., Prinja, R.K., et al. 1997, A\&A, 327, 699
\rfr Henrichs, H.F., Kaper, L., \& Zwarthoed, G.A.A. 1988, in A Decade
     of UV Astronomy with the IUE Satellite, ESA SP-281, Vol.~2, 145
\rfr Howarth, I.D., Prinja, R.K., Massa, D. 1995, ApJ, 452, L65
\rfr Kaper, L., Henrichs, H.F., Nichols, J.S., et al. 1996, A\&AS, 116, 257
\rfr Kaper, L., Henrichs, H.F., Fullerton, A.W., et al. 1997, A\&A, 327, 281
\rfr Kaper, L., Henrichs, H.F., Nichols, J.S., \& Telting, J.H. 1999, A\&A,
      344, 231
\rfr Kaufer, A., Stahl, O., Prinja, R.K., \& Witherick, D. 2006, A\&A, 447, 325
\rfr Lobel, A., \& Blomme, R. 2008, ApJ, 678, 408
\rfr Massa, D., Fullerton, A. W., Nichols, J. S., et al. 1995, ApJ, 452, L53
\rfr Morel, T., Marchenko, S.V., Pati, A.K., et al. 2004, MNRAS, 351, 552
\rfr Mullan, D.J. 1986 A\&A, 165, 157
\rfr Owocki, S.P. 1998, in Cyclical Variability in Stellar Winds,
     Eds. L. Kaper \& A.W. Fullerton, ESO Astrophysics Symposia,
     Springer-Verlag, 325
\rfr Owocki, S.P., Cranmer, S.R., \& Fullerton, A.W. 1995, ApJ, 453, L37
\rfr Prinja, R.K. 1988, MNRAS, 231, P21
\rfr Prinja, R.K., \& Howarth, I.D. 1986, ApJS, 61, 357
}

\end{document}